\documentclass[12pt]{article}
\usepackage{epsfig}
\newcommand{\beq}{\begin{equation}}
\newcommand{\eeq}{\end{equation}}
\newcommand{\beqa}{\begin{eqnarray}}
\newcommand{\eeqa}{\end{eqnarray}}
\newcommand{\beqar}{\begin{eqnarray*}}
\newcommand{\eeqar}{\end{eqnarray*}}
\newcommand{\eg}{{\it e.g.,}\ }
\newcommand{\ie}{{\it i.e.,}\ }
\newcommand{\labell}[1]{\label{#1}} 
\newcommand{\reef}[1]{(\ref{#1})}

\begin{document}

\thispagestyle{empty}

\hfill{}

\hfill{}

\hfill{CERN-TH/2001-095}

\hfill{hep-th/0104009}

\vspace{32pt}

\begin{center} 
\textbf{\Large Exact Gravitational Shockwaves and
Planckian Scattering on Branes}\\

\vspace{48pt}

Roberto Emparan\footnote{Also at Departamento de F{\'\i}sica Te\'orica,
Universidad del Pa{\'\i}s Vasco, E-48080, Bilbao, Spain.}

\vspace{12pt}

\textit{Theory Division, CERN}\\
\textit{CH-1211 Geneva 23, Switzerland}\\
\texttt{roberto.emparan@cern.ch}
\end{center}

\vspace{48pt}

\begin{abstract} 

We obtain a solution describing a gravitational shockwave propagating
along a Randall-Sundrum brane. The interest of such a solution is
twofold: on the one hand, it is the first exact solution for a localized
source on a Randall-Sundrum three-brane. On the other hand, one can use
it to study forward scattering at Planckian energies, including the
effects of the continuum of Kaluza-Klein modes. We map out the different
regimes for the scattering obtained by varying the center-of-mass energy
and the impact parameter. We also discuss exact shockwaves in ADD
scenarios with compact extra dimensions.

\end{abstract}

\setcounter{footnote}{0}

\newpage

\section{Introduction}

No matter how small the rest mass of a particle is, when it is
accelerated to energies in or above the Planck scale its gravitational
field becomes so strong that it can not be neglected. It has been known
for some time what this field looks like: a planar shock wave, whose
rays propagate parallelly to the direction of motion \cite{AS,DtH}. When
another particle crosses this wavefront, its trajectory is altered---in
other words, the second particle is scattered by the attractive
gravitational field of the Planckian-energy particle. It was shown in
\cite{tH} that the amplitude for this sort of scattering can be exactly
calculated. As it turns out, this way of computing the scattering
between the two particles corresponds to the leading approximation to
the forward scattering of two particles in quantum gravity, for
center-of-mass energy much larger than the momentum transfer
\cite{MuSo,ACV,ACV2,ACV3,vv,KaOr}.

It has been commonly assumed that, given the enormous value of the
Planck scale, Planckian energies would very hardly be attainable.
However, it has been realized in recent years that the fundamental scale
for quantum gravity may not be the usual four-dimensional Planck scale,
$M_{Pl}\sim 10^{18}$ GeV. Rather, the fundamental scale $M_*$
might be essentially anywhere between the TeV scale and $M_{Pl}$. The
latter would be a derived magnitude, adequate for describing gravity
only at low energies/large distances, and its large value would arise as
a consequence of the existence of large (ADD \cite{ADD}), or warped (RS1
\cite{RS1}), extra dimensions. If some form of scenario of low-scale
quantum gravity were actually realized, Planckian energies might be much
more accesible than previously thought. For $M_*$ in the TeV range, it
could be reached in colliders in the near future, whereas intermediate,
as well as low, scales might perhaps be probed by extreme energy cosmic
rays. Currently, the case for the latter is still open, see \eg
\cite{cosmic}, but it should be noted that the regime probed by these
cosmic rays appears to be precisely the one described in the previous
paragraph.

Given these considerations, it is natural to try to extend the analysis
of the shockwave of an ultra-high-energy particle to such scenarios with
extra dimensions. Among these, a large and particularly interesting
class regards our universe as a three-brane embedded in a higher
dimensional bulk \cite{Aka,RS0,ADD,RS1,RS2}. The focus of this paper will be
on such brane-world scenarios, and, mostly, on the Randall-Sundrum model
with an infinite extra dimension \cite{RS2}, henceforth RS2. 

The phenomenology of RS2 is not as much developed as that of ADD or RS1.
Some steps were taken in \cite{CED}. The main difference is that RS2 is
not designed to address the hierarchy problem. In fact, in RS2 the
fundamental and effective four-dimensional gravity scales are related as
$M_*=(M_{Pl}^2/\ell)^{1/3}$, and since experiment bounds $\ell$---the
curvature scale of the extra dimension---to be not larger than 1 mm, then
$M_*>10^5$ TeV, which still might be within the reach of cosmic rays. 
Nevertheless, there are variants of RS2 \cite{ADDK} with $n$ extra
dimensions which allow for much lower values of $M_*$ through
$M_*=(M_{Pl}^2/\ell^n)^{1/(n+2)}$. We will focus exclusively on RS2, but
the extension of our analysis to the models of \cite{ADDK} should not
present technical difficulties.

{}From the conceptual point of view, the RS2 model has resulted
extremely fruitful, opening up new avenues for thinking about gravity in
extra dimensions. However, the structure of the model---a three-brane in
a constant negative curvature background---has made it very difficult to
analyze gravity on it in an exact way. It is particularly important to
know what is the gravitational field created by sources localized to the
brane. So far, the only known exact solutions, constructed in
\cite{EHM}, describe black holes in a lower-dimensional setting---a
two-brane in a four-dimensional bulk. Hence, the construction of other
simple, exact solutions in this model is of obvious interest. A main
part of this paper (Section \ref{thesoln}) is devoted to constructing
the exact gravitational shockwave of an effectively massless particle
within the RS2 model. To our knowledge, these are the first exact
solutions to describe the gravitational field of a localized source on a
RS brane in $AdS_5$ (or higher dimensions). With this solution in hand,
we will follow \cite{tH} and \cite{ACV} in Section \ref{scat} to describe
certain aspects of Planckian scattering on the brane.

Finally, given their phenomenological interest, one would also like to
have a similarly exact description of shockwaves in ADD scenarios. If the
gravitational backreaction of the brane is neglected (as it usually is,
but see \cite{CEG}), this turns out to be much easier than in the RS2
model. Therefore, we will present these solutions in Section
\ref{addshock}.

Throughout this paper we will denote the conventional (four-dimensional)
Planck mass as $M_{Pl}\equiv G_4^{-1/2}$, while $M_*=G_5^{-1/3}$ will be
the fundamental (five-dimensional) mass. What is precisely meant by
`Planckian energy', and in which regime, will be discussed in Section
\ref{scat}. We also take $\ell$ to be larger than the fundamental (or
string) scale. This seems reasonable, since otherwise the semiclassical
description of the RS setup using Einstein gravity would not be
reliable.

\section{Gravitational Shockwave on the
RS Brane}\label{thesoln}

Working in an arbitrary number of dimensions, the RS2 scenario describes
a $(d-1)$-brane in the $AdS_{d+1}$ spacetime, the case of most relevance
being obviously $d=4$\footnote{There is only a single brane here, so this
is different from the higher-dimensional scenarios of \cite{ADDK}.}. The
ground state metric is
\beq
ds^2=dy^2+e^{-2|y|/\ell}\eta_{\mu\nu}dx^\mu dx^\nu\,,
\eeq
with $\mu,\nu=0,\dots,d-1$. The coordinate $y$ measures the proper
distance transverse
to the brane, which is itself located at the orbifold point $y=0$. It
is at times also convenient to use another form for the metric, by
changing the bulk coordinate to
\beq
z=\ell(e^{y/\ell}-1)\,,
\eeq 
so
\beq
ds^2={\ell^2\over (\ell+|z|)^2}(dz^2+\eta_{\mu\nu}dx^\mu dx^\nu)\,.
\labell{zcoord}\eeq
The brane is now at $z=0$.

Our starting point is a particle at rest on the brane. In Ref.\
\cite{AS}, Aichelburg and Sexl showed that, in four flat dimensions, the
metric for the shockwave could be constructed by performing a boost to
the speed of light on the Schwarzschild solution. In the present case,
exact solutions for black holes on RS branes in $AdS_{d+1}$ are unknown
except for the low-dimensional model in $d=3$ \cite{EHM}. We will instead
use the approximations that have been constructed up to linearized order.
As in the case of \cite{AS}, performing an ultrarelativistic boost will
have the effect that only the linearized part of the solutions remains
important. 

Therefore, let us place a source on the brane, localized on it, which
means that its stress-energy tensor $t_{\mu\nu}(x)$ has components only
along the brane-world indices, and that it depends solely on the
brane-world coordinates. The equations for the linearized perturbation
$\eta_{\mu\nu}+h_{\mu\nu}$ induced by the source have been the subject
of a number of papers, including \cite{RS2,GT,gkr,aimvv}. The final result
can be given in terms of Fourier transforms with respect to the
brane-world coordinates,
\beq
h_{\mu\nu}(q,y)=\int d^4x \;e^{-iq_\sigma x^\sigma} h_{\mu\nu}(x,y)\,,
\eeq
in the form \cite{aimvv}
\beq
\tilde h_{\mu\nu}(q,y)=8\pi G_{d+1}\left[t_{\mu\nu}(q)-{1\over
d-1}\left(\eta_{\mu\nu}-{q_\mu q_\nu\over q^2}\right)t\right]
e^{d|y|\over
2\ell}{K_{d/2}(e^{|y|/\ell}\ell q)\over q
K_{d/2-1}(\ell
q)}\,.
\labell{linsol}\eeq
The tilde denotes the tracefree perturbation $\tilde
h_{\mu\nu}=h_{\mu\nu}-{1\over d}\eta_{\mu\nu} h$, and the solution is
expressed in terms of Bessel $K_\nu$ functions. Also,
$G_{d+1}=M_*^{-d+1}$ is the $d+1$ dimensional gravitational
constant. The trace of the perturbation must satisfy 
\beq
h|_{y=0}=-{32\pi G_{d+1}\over (d-1)\ell q^2}t\,,
\eeq
but in fact we will not need it.

For a point particle at rest, of mass $m$, the stress-energy tensor is
$t_{00}(q)=2\pi m\delta(q_0)$. The corresponding metric perturbation
can be readily found from the above formulas, even if the inverse
Fourier transforms can only be explicitly evaluated in certain limits.
Nevertheless, we can still boost the solution in Fourier
space.
When boosted to
high energies the particle becomes ultrarelativistic, and then we can
effectively take $v\to 1$, while keeping the\ momentum $p=\gamma mv$
fixed. Instead of boosting the solution $h_{\mu\nu}$ for a particle at
rest, we will, equivalently, find the solution that corresponds to the
stress-energy tensor of a boosted particle. This stress-energy tensor
transforms under the boost, and then as $v\to 1$, as
\beqa
t_{00}(q)&=&2\pi\gamma m\; \delta(q_0+vq_1)\to 2\pi p
\;\delta(q_0+q_1)\,,\nonumber\\
t_{01}(q)&=&v t_{00}(q)\to t_{00}(q)\,,\\
t_{11}(q)&=&v^2 t_{00}(q)\to t_{00}(q)\nonumber\,,
\labell{masslessstr}
\eeqa
which is effectively the stress-energy tensor of a massless particle. We
can now plug this into equation \reef{linsol} to obtain the desired form
of the solution. It is important to note that the stress-tensor
\reef{masslessstr} is trace-free. Hence the metric perturbation
$h_{\mu\nu}$ can be taken to be trace-free too. This implies that the
so-called ``brane-bending'' effect \cite{GT} is absent. The gravitons
with polarizations transverse to the brane are not excited and hence the
brane does not bend into the bulk. 

Given that we are dealing with a limiting lightlike source, it is
convenient to work with the light-cone coordinates $u=t-x$, $v=t+x$. In
terms of these, the perturbed metric for a null source takes the form
\beq
ds^2=dy^2+e^{-2|y|/\ell}(-dudv+dx^i dx^i +h_{uu}(u,x^i,y)du^2)\,,
\labell{ppwave}\eeq
where $i=2,\dots,d-1$ labels the coordinates in the brane-directions
transverse to the propagation. Plugging the stress-energy tensor
\beq
t_{uu}=2\pi p\;\delta(q_0+q_1)\,,
\labell{sefour}\eeq
into \reef{linsol}, and transforming back to coordinate space we get
\beq
h_{uu}(u,x^i,y)={4G_{d+1}\over (2\pi)^{d-3}}p\;\delta(u)e^{d|y|\over
2\ell}\int d^{d-2}q_i\; e^{i q_i x^i}{K_{d/2}(e^{|y|/\ell}\ell
q)\over q K_{d/2-1}(\ell
q)}\,,
\labell{exactsol1}\eeq
where now $q=|q_i q_i|^{1/2}$ is the modulus of the projection of $q_\mu$
on the plane transverse to the propagation of the particle, \ie parallel
to the wavefront. The Fourier transforms cannot, for $d\geq 4$, be
carried out fully explicitly, but at least the angular integrations that
appear can be performed,
\beq
h_{uu}(u,r,y)={4G_{d+1}\over (2\pi)^{d-4\over
2}}p\;\delta(u){e^{d|y|\over
2\ell}\over r^{d-4\over 2}}\int_0^\infty dq {q}^{d-4\over 2}
J_{d-4\over 2}(q r){K_{d/2}(e^{|y|/\ell}\ell q)\over 
K_{d/2-1}(\ell
q)}\,,
\labell{exactsol2}
\eeq
where $r$ is the radial distance on the wavefront on the
brane, transverse to the direction of propagation of the particle. 
Note that away from the wavefront, the perturbation vanishes.

This solution is in fact an exact one: for a $d+1$-dimensional metric
of the plane wave
form \reef{ppwave}
the exact Einstein tensor is
\beqa
G_{yy}&=&{d(d-1)\over 2\ell^2}g_{yy},\nonumber\\
G_{\mu\nu}&=&\left({d(d-1)\over 2\ell^2}-{2(d-
1)\over\ell}\delta(y)\right)g_{\mu\nu}-{1\over 2}
\partial_\mu u\partial_\nu u\left[
e^{-2|y|/\ell}\left(\partial^2_y -{d\over
\ell}\partial_y\right)+\nabla^2_x\right]h_{uu}\,.
\labell{einstensor}
\eeqa
All other components vanish. This exact Einstein tensor is linear in
$h_{uu}$. Hence by solving the equations at linearized order we have
actually solved them to all orders. This linearization, which had been
noted earlier in \cite{CG}, allows to construct exact plane waves
localized on the brane in the RS model. 

Therefore, the solutions \reef{exactsol2} provide an exact
description of the gravitational field of a lightlike point source
localized on the brane.

Let us now focus on the case of $d=4$, and in particular, on the metric
at the location of the brane, $y=0$. Although we have not been able to
perform the last integration in \reef{exactsol2} explicitly, we can
approximate it in several limits. At large distances from the source on
the wavefront on the brane, $r\gg \ell$, we can expand the Bessel
functions for small $q$, to find
\beq
h_{uu}=-4 G_4 p\;\delta(u)\left[ \log(r^2/
\ell^2)
-{\ell^2\over r^2}+{2\ell^4\over r^4}\left(\log
(r^2/
\ell^2)-1\right)+\dots\right]\,.
\labell{longdist}\eeq
This result has been written already in terms of the effective
gravitational coupling constant induced on the brane,
$G_4=G_5/\ell$. As was the case for static point masses,
the first correction, $\sim -\ell^2/r^2$, does not resemble the profile
of a five-dimensional shockwave (which would go like $\ell/r$), rather
that of a six-dimensional one. However, at short distances ($r^2+y^2\ll
\ell^2$), instead, it is easy to see that the five dimensional form of
the shockwave is recovered, due to dominance of KK modes. 
More explicitly, on the brane at $r\ll\ell$,
\beq
h_{uu}=-4 G_4 p\;\delta(u)\left[-{\ell\over r}+{3\over 2}\log
(r/\ell)+{3r\over 8\ell}+\dots\right]\,.
\labell{shortdist}\eeq

A different form for the solution, which is better suited for numerical
evaluation of the integrals, can be obtained by applying the
analysis in \cite{GT} to the source of \reef{sefour}:
\beqa
&h_{uu}(u,r,y)=&-4 G_4 p\;\delta(u)\biggl[e^{-2|y|/\ell} 
\log (r^2/\ell^2)\nonumber\\
&&-
{2\ell\over\pi}\int_0^\infty dm\;K_0(m r){Y_1(m\ell)J_2(m\ell
e^{|y|/\ell})-J_1(m\ell)Y_2(m\ell
e^{|y|/\ell})\over
J_1^2(m\ell)+Y_1^2(m\ell)}\biggr]\, .
\labell{otherform}\eeqa
The zero mode term has been split from the continuum of Kaluza-Klein
modes of mass $m$.
Again, this is an exact form for the solution. The factor
$e^{-2|y|/\ell}$ indicates the suppression of the solution into the
bulk. On the brane the solution becomes
\beq
h_{uu}(u,r,y=0)=-4 G_4 p\;\delta(u)\left[ \log (r^2/\ell^2)-
{4\over\pi^2}\int_0^\infty {dm\over m}{K_0(m r)\over
J_1^2(m\ell)+Y_1^2(m\ell)}\right]\, .
\labell{kkdecomp}\eeq

\begin{figure}[t]
\centerline{\epsfig{file=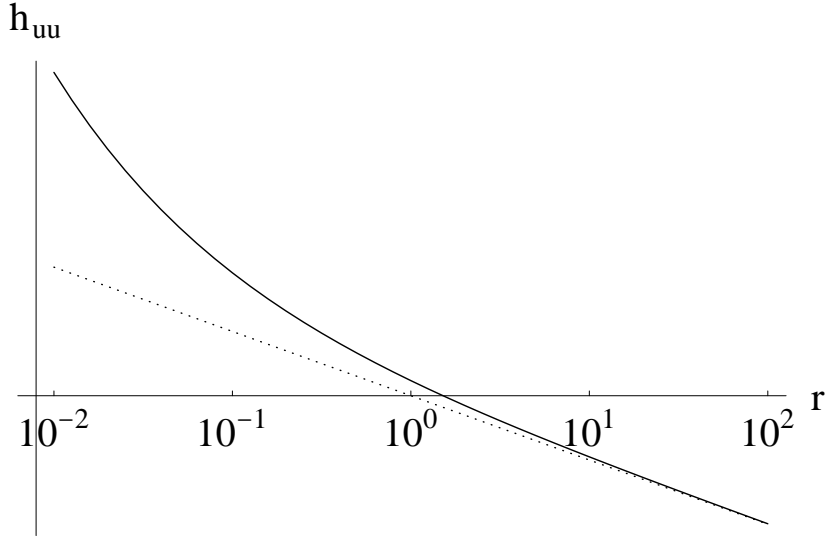,width=11cm}}
\vskip 5mm
\caption{Log-scale plot of the profile of the shockwave
$h_{uu}(r)$ on the RS three-brane. It clearly exhibits the
deviation, due to the Kaluza-Klein modes, from
the four-dimensional
logarithmic solution (dotted line). The
units for $r$ are such that
$\ell=1$.}
\label{logwave}
\end{figure}

\begin{figure}[t]
\centerline{\epsfig{file=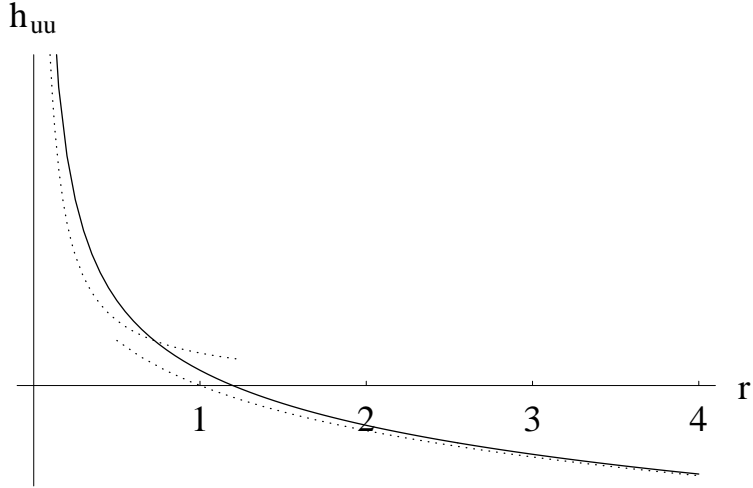,width=10cm}}
\vskip 5mm
\caption{The shockwave profile $h_{uu}(r)$ on an RS three-brane (solid
line). It interpolates between the leading order behaviors at short
distance ($\sim r^{-1}$) and large distance ($\sim -\log r$),
in dotted lines.
Again, we have set
$\ell=1$.}
\label{waveplot}
\end{figure}

We have used this latter form of the solution to plot $h_{uu}(r)$ in
Fig.~\ref{logwave}. The figure very clearly shows how the Kaluza-Klein
modes introduce, at distances $r< \ell$, an enhancement of the
gravitational shockwave relative to the zero-mode truncation, \ie the
leading log term in \reef{longdist} and \reef{kkdecomp}. In
Fig.~\ref{waveplot} we exhibit how the exact solution interpolates
between the four-dimensional behavior at large distances, and
five-dimensional gravity at short distances. In the latter case, it is
interesting to note that the leading order approximation, $1/r$, yields a
weaker effect than the exact value. The first correction in
\reef{shortdist}, $\sim -3\log(r/\ell)/2$, becomes in fact of some
importance.

\subsection*{The shockwave on a two-brane}

It is interesting to consider separately the low-dimensional case of
$d=3$, corresponding to a domain wall in $AdS_4$. For the shock wave
solution all the calculations can be carried out explicitly to
the end, since the Bessel functions involved can be finitely expressed
in terms of elementary functions. Using the form of the metric in terms
of the $z$ coordinate of \reef{zcoord} we find
\beq
h_{uu}=-4 G_4 p\;\delta(u)\left(\log(r^2+z^2)+{2|z|\over
\ell}+{2r\over
\ell}\arctan{r\over |z|}\right)\,.
\labell{asontwo}
\eeq
Along the brane
at $z=0$ this reduces to
\beq
h_{uu}(u,r,z=0)=- 8G_3 p\;\delta(u)\left(\pi |r|+\ell\log(r^2)\right)\,,
\label{onthebrane}\eeq
where we have used $G_4=2\ell G_3$. As explained above, the linearized
solution is in fact an exact one. As was the case for black holes on a
two-brane constructed in \cite{EHM}, the exact metric on the brane
\reef{onthebrane} is precisely the sum of the $2+1$ dimensional
($\propto |r|$) and $3+1$ dimensional ($\propto \log(r^2)$)
solutions. Observe that in the bulk of spacetime, the
four-dimensional form of the solution ($\sim \log(r^2+z^2)$) is recovered
for small $r$ and $z$.

In \cite{DtH} it was shown how the Aichelburg-Sexl solution can be
constructed by a cut-and-paste method performed in flat Minkowski space.
It would be interesting to show how this exact solution is obtained by
similarly cutting and pasting patches of $AdS_4$ (see \cite{PG} for
possibly related work).

\section{Planckian Scattering on the Brane}\label{scat}

We will now study the elastic forward scattering amplitude, in a regime
where the center-of-mass energy is at a very large scale, and is much
larger than the momentum transfer, $s/|t|\gg 1$. Gravitons are expected
to dominate over other interactions above the Planck energy. Obviously,
one must specify what is meant by `Planckian energies' here, \ie whether
$E>M_*$ or $E>M_{Pl}=M_*\sqrt{M_*\ell}\gg M_*$. Recall that the
assumption of `Planckian center-of-mass energies' has several
motivations. First, it ensures that the rest mass $m_0$ of the particle
is negligible. To this effect, we just need energies $E\gg m_0$, but not
necessarily Planckian. More importantly, at energies above the Planck
scale the effective dimensionless coupling $\alpha_G\equiv s/M_{Pl}^2$
becomes large and gravity is expected to be strongly coupled.
Furthermore, due to the growth of this coupling with energy, it will
dominate over any other interactions. In the present case, however, one
should consider first the distance scale that is being probed. If the
impact parameter $b$ is much larger than $\ell$, then the graviton
zero-mode dominates over KK modes. In this regime, Planckian energy will
necessarily mean $E>M_{Pl}$. The graviton zero-mode will then be strongly
coupled. Instead, for $b<\ell$ the KK modes dominate and the interaction
becomes five-dimensional. Here our methods can also be applied to the
regime of $M_*<E<M_{Pl}$, but this will not necessarily imply that the KK
modes are strongly coupled---we will discuss when they are. Gravity need
not dominate over other interactions in this regime. But for $E>M_{Pl}$
five-dimensional gravity will always be strongly coupled\footnote{This is
not surprising, since individual KK modes couple with constant
$G_4=M_{Pl}^{-2}$.}.

The forward scattering of two particles (or strings) at Planckian
energies has been studied in the past \cite{tH,MuSo,ACV}, however, the
possibility of new dimensions opening up was not generally considered.
Some discussion of this point has been given in \cite{BaFi}. Our analysis
is somewhat complementary to that in \cite{BaFi}, but we will go into
more detail at several points. From the technical point of view, we
mainly build up on the work of \cite{tH} and \cite{ACV,ACV2}. The regime
of Planckian energies and large $s/|t|$ can be treated in the eikonal
approximation---a resummation of an infinite number of graviton ladder
and cross-ladder diagrams, which dominates the elastic forward
scattering. Although it resums contributions from all orders in the
coupling constant, this approximation does not actually probe quantum
gravity effects. Effectively, graviton loops are suppressed if the impact
parameter $b$ is much larger than the fundamental length $M_*^{-1}$ (the
momentum exchanged by each graviton is much less than $M_*$). Hence,
corrections in $1/(b M_*)^2$ are neglected. Notice also that
four-dimensional graviton loops are suppressed by the much larger factor
$1/(b M_{Pl})^2$. 

Another important point is the possibility of black hole physics entering
at impact parameters of the order of the Schwarzschild radius associated
to a given center-of-mass energy. Precise calculations are way beyond any
computational scheme available (it involves tree-level graviton exchange
to all orders, and possibly beyond perturbation theory), but we will
discuss this later in mostly qualitative terms.

\begin{figure}[th]
\centerline{\epsfig{file=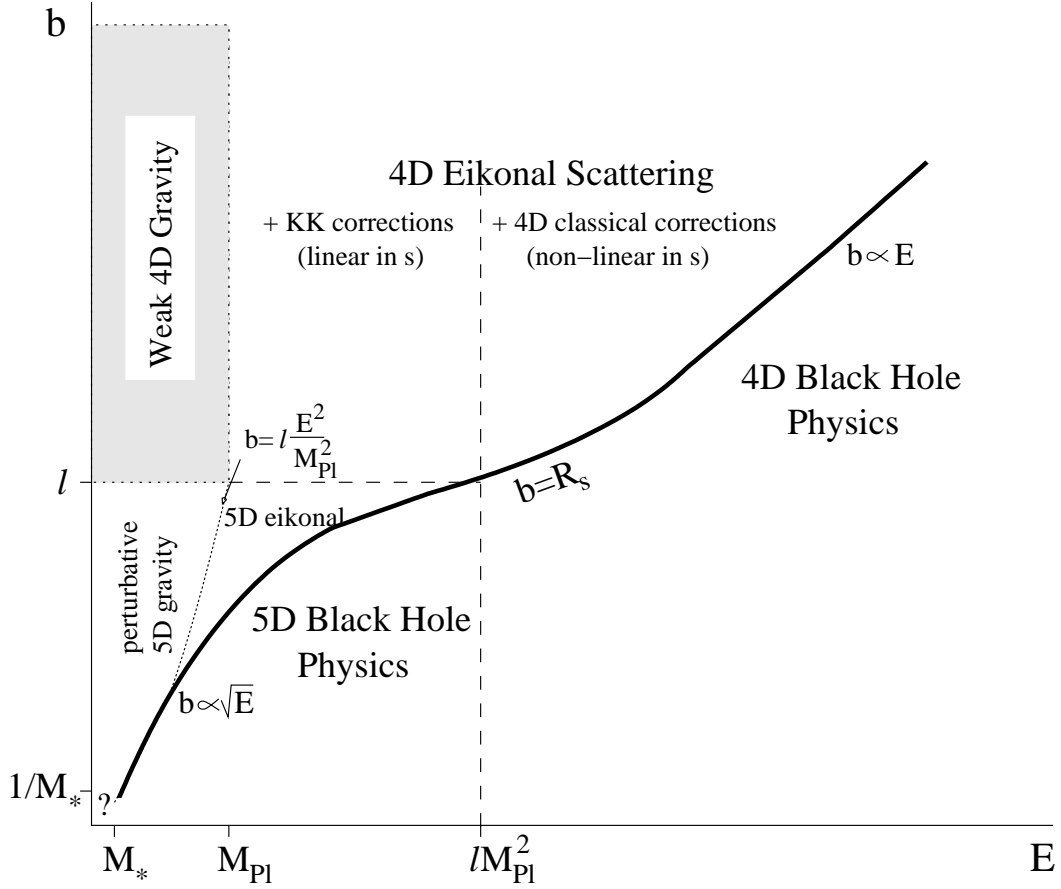,width=14cm}}
\vskip 5mm
\caption{The different regimes for the scattering at energies $E>M_*$
and impact parameters $b>1/M_*$. The boundaries between regimes are
merely indications of where the crossover from one behavior
to another takes place.}
\label{bvsE}
\end{figure}

Following \cite{tH} (see also \cite{AK}) the shockwave geometry directly
yields the relevant information needed to compute the eikonalized
gravitational scattering amplitude for two particles at large center of
mass energy $\sqrt{s}=2E$. Let $a(s,b)$ be the scattering amplitude
expressed as a function of the
impact parameter $b$. In eikonal form
\beq
a(s,b)=e^{2i\delta(s,b)}\,.
\labell{asb}\eeq
If we identify the impact parameter $b$ with the distance in the
direction transverse to the propagation of the
shockwave,
$b=\sqrt{x^ix^i}$, then $\delta(s,b)$ can be read off from the shockwave
metric as
\beq
h_{uu}=8 p\;\delta(u){\delta(s,b)\over s} \,.
\labell{hdelta}\eeq

Having $a(s,b)$, one can compute the amplitude for a given momentum
transfer, $t=-q_iq_i=-|q|^2$, by transforming
\beqa
{1\over s}a(s,t)&=&{2\over i}\int d^2 x\; e^{iq_i x^i}\left(e^{2i
\delta(s,b)}-1\right)\nonumber\\
&=&{4\pi\over i}\int_0^\infty db\; b J_0(b q) e^{2i\delta(s,b)}\,.
\labell{ast}\eeqa 
It is important to notice here that the Fourier transform is a
two-dimensional one, even if the shockwave front is three-dimensional.
The reason is that we are considering the scattering of particles
confined to the brane, and therefore the impact parameter is restricted
to the two transverse directions along the brane. The available phase
space is reduced in comparison to particles that propagate freely in the
bulk. As a consequence, even at short distances $b\ll \ell$ the
scattering amplitude $a(s,t)$ will differ from the ``really
five-dimensional'' one.

Let us first consider impact parameters $b$ larger than $\ell$. In this
regime, which is essentially four-dimensional, the eikonal approach is
only justified if $E>M_{Pl}$, and not for $M_*<E<M_{Pl}$. At $b>\ell$
and energies below $M_{Pl}$, the gravitational interaction is very weak
and likely to be dominated by other interactions. But for $E>M_{Pl}$
gravitons dominate and we can obtain the eikonal from \reef{longdist}
and
\reef{hdelta}. Keeping only the two first terms, we get
\beq
\delta(s,b)=-G_4 s\left(\log (b/\ell)-{\ell^2\over 2 b^2}\right)\,.
\labell{4deik}\eeq
The KK correction to the leading logarithmic term,
\beq
\delta_{KK}(s,b)=G_4 {s\ell^2\over 2 b^2}\,,
\eeq
grows linearly with $s$, just like the four-dimensional term, a fact
that distinguishes it from other non-linear corrections. The expansion
parameter for KK corrections is $\ell^2/b^2$. Classical
corrections to the eikonal, that include the graviton self-interaction
vertices, but still at graviton tree level, have the expansion parameter
$G_4^2 s/b^2=s/(M_{Pl}^2 b^2)$ \cite{ACV2}. Since we are assuming that
$M_{Pl}\gg \ell^{-1}$, then the KK corrections will be larger than these
4D classical corrections up
to energies $E\sim M_{Pl}^2 \ell$. 

With \reef{4deik}, the integral \reef{ast} can be evaluated at a saddle
point $b=b_s$ such
that
\beq
q=-2{\partial\delta(s,b)\over \partial b}|_{b_s}\nonumber\\
\simeq {2 G_4 s\over b_s}\left(1+{\ell^2\over b_s^2}\right)\,.
\eeq
As long as the saddle point satisfies $b_s>\ell$, it is justified to
ignore the physics at smaller $b$ in the integral \reef{ast}. Notice
that the momentum exchanged at a given impact parameter is larger than
in four dimensions, due to the exchange of KK modes. Equivalently, the
deflected angle,
\beq
\theta\simeq 2\sqrt{-t\over s}\simeq {4 G_4 E\over
b_s}\left(1+{\ell^2\over b_s^2}\right)\,,
\eeq
is increased, showing the extra attraction that KK modes induce.

Let us now move to the short distance regime $b<\ell$. In this case,
keeping just the leading order from \reef{shortdist} one gets
\beq
\delta(s,b)\simeq {G_4 s \ell\over 2 b}\, .
\labell{shortd}\eeq
This eikonal phase is small if $2 b>G_4 s\ell=s/M_*^3$, and therefore leads
to a perturbative regime, which, for the fixed $t$ amplitude, is at
momentum transfer $q<2/(G_4 s\ell)$. This is, the Born term dominates the
expansion, and one can do without the eikonal resummation. This is in
contrast to the previous situation, where the amplitude was always
non-perturbative, and dominated by a saddle point. Gravity here is
five-dimensional (it involves all the KK modes), and the interaction is
stronger than it would be in a four dimensional setup. But it is not
strongly coupled. 

Starting at fixed energy $M_*<E<M_{Pl}$, one enters this regime when the
impact parameter gets below $\ell$. The fixed-$t$ amplitude becomes 
\beq
a(s,t)\simeq 8\pi s\int_0^\infty db\; b J_0(bq)\delta(s,b)= 4\pi
G_4\ell\; {s^2\over q}\,.
\eeq
This is different from the usual perturbative result for gravity, which
is (in any dimension) $\sim G_n s^2/q^2$. The reason has been explained
above: although the interaction is five dimensional, the scattered
particles are confined to the four-dimensional brane. 

For $q>2/(G_4 \ell s)$ we enter a strong coupling regime and the
amplitude $a(s,t)$ is again dominated by a saddle point, this time at
\beq
q\simeq{G_4 s\ell\over b_s^2}\,.
\eeq
In this case the full eikonal resummation is relevant, and the amplitude
evaluated at the saddle point is
\beq
a(s,t)\simeq 4\pi\sqrt{G_4\ell\over 2}\left({s\over
q}\right)^{3/2}e^{2i\sqrt{G_4\ell s q}-i\pi/2}\,.
\eeq
The non-analytic dependence on the coupling shows the non-perturbative
character of the amplitude.

Let us note that the 't Hooft poles \cite{tH} do not appear in these
amplitudes. When the eikonal phase is purely logarithmic (\ie 4D), these
poles arise from the $b\to 0$ region in the integral \reef{ast}. Here,
however, the eikonal changes from 4D behavior to 5D behavior before
getting to $b\to 0$, and the poles disappear\footnote{For the eikonal
phase \reef{shortd}
one can actually compute exactly the amplitude, $a(s,t)=8\pi G_4\ell 
{s^2\over q}J_1(e^{-i\pi/4}\sqrt{2G_4s \ell
q})K_1(e^{-i\pi/4}\sqrt{2G_4s \ell q})$. Here one sees explicitly that
the 't Hooft poles are absent.}. Since the 't Hooft poles could be
interpreted as a remnant of the bound states in the 4D Coulomb potential
\cite{ACV3,KaOr,Dit}, it is no surprise that they are absent here, since
the 5D Coulomb potential does not have (stable) bound states.

The results above cease to be reliable when the impact parameter becomes
of the order of, or smaller than the gravitational (Schwarzschild)
radius, $R_s$. Indeed, for $b\ll R_s$ one expects gravitational collapse
to take place. The details, though, are expected to be very complicated,
particularly for intermediate scales $b\sim R_s$. Although the full
scattering problem is way beyond the techniques used here (see
\cite{DEPa}), one can assume this regime will be dominated by black hole
physics. Hence the discussion will be at a qualitative level. Additional
discussion of related issues can be found in \cite{gk}.

In a scenario like this, the Schwarzschild radius $R_s$ changes depending
on the regime one is in. In the effective four dimensional regime of
distances larger than $\ell$, the classical gravitational radius is
\beq
R_s\simeq 2G_4 E\,.
\labell{largebh}\eeq
The black hole is a `pancake' in this regime, with a very
small extent into the bulk $\sim \ell\log(R_s/\ell)\ll R_s$
\cite{EHM,gkr}. The physics of these black holes
is described by four-dimensional laws.
Instead, at distances shorter than $\ell$,
\beq
R_s\simeq \sqrt{8 G_5 E\over 3\pi}=\sqrt{8 G_4\ell E\over 3\pi}\, .
\labell{smallbh}\eeq
These small black holes are roughly spherical in five dimensions. The growth
with $E$ changes from one regime to the other, with some smooth
interpolation at distances $\sim\ell$.

The total cross section for producing these black holes can be estimated
to be of the order of the corresponding black hole area. Depending on
whether the black hole is a large or a small one, we have
\beqa
\sigma&\sim& {s\over M_{Pl}^4}\qquad \mathrm{for}\ E>\ell
M_{Pl}^2\,,\nonumber\\
\sigma&\sim& {\sqrt{s}\over M_*^3}={\ell \sqrt{s}\over M_{Pl}^2}\qquad
\mathrm{for}\ E<\ell
M_{Pl}^2\,.
\eeqa
Since the particles scatter on the brane, the relevant magnitude for
producing a small black hole is not the five-dimensional black hole area
(which is in fact a volume), but rather its section along the brane,
which can be assumed to be along an equator of the horizon. Notice also
that an effectively four-dimensional black hole will not be formed until
$E>M_{Pl}^2\ell\gg M_{Pl}$.

The black holes thus created will evaporate by emission of Hawking
radiation. In either regime (large or small), the radiation will be
emitted mostly along the brane \cite{EHM,EHM3}.

The different regimes in the $(E,b)$ plane are displayed in Figure
\ref{bvsE}. The region marked ``weak 4D gravity" is one where
four-dimensional gravity is weakly coupled and the interaction dominated
by single graviton exchange, which we have not discussed here---the
leading amplitude is the same as the eikonal, up to a phase. The regions
labelled ``eikonal" are ones where gravity is strongly coupled, and the
full eikonal resummation of the amplitude is needed. The amplitudes are
non-perturbative there. The curve $b=R_s$ is an interpolation between
\reef{largebh} and \reef{smallbh}. Note that the scattering is directly
sensitive to the extra dimensions only at energies below $\ell M_{Pl}^2$.
Going to higher energies does not actually lead into five-dimensional
physics.

The picture should be valid down to impact parameters $b>1/M_*$ (or the
string length), where new physics takes over. This regime is beyond the
techniques used here.

\section{Exact Shockwaves in the ADD scenario}\label{addshock}

The construction of exact shockwaves in ADD scenarios is simpler than in
RS. We discuss it briefly here.

The ADD scenario consists of a three-brane (admitting a
Poincar\'e-invariant vacuum) living in a $4+n$-dimensional spacetime. In
the most basic setup, the bulk is empty and the gravitational
backreaction of the brane is neglected: the brane is simply a $3+1$
hypersurface embedded in the bulk. The extra dimensions are supposed to
be compactified on a certain manifold $\mathcal{M}$. If the bulk is
empty, then the metric on $\mathcal{M}$ has to be Ricci-flat. Hence, if
we label the brane coordinates by $x^\mu$, and the
transverse coordinates by $y^a$, the vacuum is
\beq
ds^2=\eta_{\mu\nu}dx^\mu dx^\nu +\hat g_{ab}\;dy^a dy^b\,,
\eeq
where $\hat g_{ab}$ is the metric on $\mathcal{M}$, and the brane is at
a certain point in $\mathcal{M}$, say, at $y^a=0$.

The linearization of the Einstein equations that occurs for the
solutions we seek simplifies again the construction.
A plane-fronted wave will be of the form
\beq
ds^2=-dudv+dx^i dx^i+h_{uu}(x,y)du^2+\hat g_{ab}dy^a dy^b
\eeq
($i=2,3$).
For a lightlike source localized on the brane the Einstein equations
become
\beq
(\nabla^2_{x}+\nabla^2_{y})h_{uu}(x,y)=-16\pi G_{4+n}
t_{uu}(x)\delta^{(n)}(y)\,,
\eeq
where we have split the Laplacian operator in the wavefront into the
brane $\nabla^2_{x}$ and the bulk $\nabla^2_{y}$ parts. The problem is
now a rather standard one. As in the RS case, a way to solve this
equation is by first Fourier-transforming the brane
coordinates,
\beq
(-q^2+\nabla^2_{y})h_{uu}(q,y)=-16\pi G_{4+n}
t_{uu}(q)\delta^{(n)}(y)\,,
\eeq
One now needs the (massive Euclidean) Green's function in the transverse
bulk
space,
\beq
(-q^2+\nabla^2_{y})G(q,y)=\delta^{(n)}(y)\,.
\labell{massg}\eeq
For the null pointlike source \reef{sefour}, the solution is then
\beq
h_{uu}(u,x^i,y)=-16\pi G_{4+n} p\;\delta(u)\int {d^2 q\over (2\pi)^2}
e^{iq_ix^i} G(q,y)\,,
\labell{addsoln}
\eeq 
which is the analogue of \reef{exactsol2}. Obviously, one can as well
give the solution as an analogue of \reef{otherform} by finding the
eigenfunctions of the operator in \reef{massg}.
As remarked above, it is the linearized character of the equations which
allows to perform the entire construction. The main problem lies in
calculating the Green's function \reef{massg} in the extra space. 

For an illustration, consider the case where the extra dimensions are
compactified on a torus $T^n$. Then, instead of \reef{addsoln}, the
solutions are most easily  obtained by using the method of images, \ie by
constructing a periodic array of $4+n$ dimensional shockwaves. Since the
equations are linear, one simply superimposes the individual solutions.
If, for simplicity, the torus is a square one, $y^a\sim y^a+L$,
then standard manipulations yield
\beq
h_{uu}(u,r,y)=-4G_4
p\;\delta(u)\left[\log(r^2)-2{\sum_{n_a\in\mathbf{Z}^n}}' 
K_0(m_n r)e^{-i n_a y^a/L}\right]\,,
\labell{solnadd}\eeq
where the sum is over vectors $n_a$ on a square lattice excluding the
origin (the zero mode has been split already), and $n=(n_a n_a)^{1/2}$
yields the mass of the Kaluza-Klein modes $m_n=n/L$. Recall that
$K_0(m_n r)$ is the Yukawa potential in two dimensions (\ie on the
wavefront on the brane). The solution is an exact one.

One can repeat the analysis of Planckian scattering performed in the
previous section. Details may change (\eg the classical gravitational
radius at short distances scales as $R_s\sim E^{1/(n+1)}$) but the
qualitative features should be similar.

\section{Concluding remarks}

In this paper we have presented solutions for gravitational shockwaves
propagating along branes, \reef{exactsol1}, \reef{otherform},
\reef{solnadd}, and argued they are in fact exact solutions, reduced to a
single quadrature or series. For the case of the RS2 model on a
two-brane, the solutions admit a simple explicit form \reef{asontwo}. 

In the past, gravitational shockwaves have been a useful tool for
studying extreme effects in quantum gravity. As such, besides the
studies of forward scattering at Planckian energies, they have also been
studied within the AdS/CFT correspondence \cite{IH}. In fact, the
context in the latter case is somewhat related to the one in this paper.
In both cases the shockwaves propagate in an $AdS_5$ spacetime. However,
it appears the solutions considered in \cite{IH}, where the wave
propagates into the bulk of $AdS_5$, are different from the ones we
discuss here, which propagate along a brane at a fixed radius from the
`center' of the $AdS_5$ space. Shockwaves in curved spacetimes and
higher dimensions have also been studied earlier in \cite{sfetsos}, in
particular there is some overlap with our elementary discussion in
Section~\ref{addshock}.\footnote{Other work in the string context can be
found in \cite{das}.}

The shockwaves on the brane may be thought of as the limiting cases of
black holes on the brane when infinitely boosted, even though such black
hole solutions remain unknown for $n>2$. However, there is a significant
difference between shockwaves and black holes in these brane-world
models. For black holes of a given mass $M$ on an RS brane there are two
different regimes, which could be called the ``large black hole'' (or
``black pancake'') and the ``small black hole'' regimes, depending on
whether, roughly, $M >\ell$ or $M<\ell$, respectively. These two regimes
could be clearly distinguished in the exact solutions constructed in
\cite{EHM}, and we have discussed some aspects in Section \ref{scat}.
There is no such distinction for the shockwaves: the description is the
same whether $p$ is large or small, and the shape of the solution only
gets rescaled by changing $p$. This is a consequence of the linearity of
the solution, which implies a simple linear dependence on $p$.

For black holes in the compact spaces relevant to ADD, there are also two
different phases according to whether the horizon radius is smaller or
larger than $L$. Small black holes are localized on the brane, whereas
large black holes are black strings which are translationally invariant
(hence delocalized) along the extra dimensions (no pancakes here). The
Gregory-Laflamme instability \cite{gl} separates the two phases. In
contrast, the shockwaves are always localized. As the energy of the
shockwave is changed, the solution simply scales linearly with $p$ and
there appears to be no reason why it should delocalize. Notice the
solution \reef{solnadd} is an exact one, whereas for black holes the
exact localized solutions in a compact space are unknown. ``Shockwave
strings'' which are translationally invariant along the extra dimensions
can be constructed, but they require translationally invariant sources
and do not seem to be relevant here\footnote{Such ``string shockwave''
solutions can also be constructed for the RS2 model, but in this case
they are even more unphysical due to their strong singularity at the AdS
horizon.}. In fact, the shockwave strings are marginally stable to
perturbations of the Gregory-Laflamme type. The absence of an instability
is not surprising if one considers that shockwaves possess no horizons
and hence no entropy. Thermodynamical arguments play no role here.

Regarding Planckian scattering on the brane, we have mapped out a
considerable portion of the different regimes that should
be amenable to a semiclassical analysis. For $b<\ell$, the expressions
obtained for the fixed $t$ amplitude account for the fact that the
interaction between the particles is five-dimensional, but the particles
themselves move only in four dimensions.

We have not discussed string effects. It is not clear whether the
results of \cite{ACV} in a flat space can be applied to this setting
even at distances much shorter than $\ell$, where the curvature effects
of $AdS_5$ would be negligible. One would first need a concrete
embedding of RS2 in string theory, and even then, solving string theory
in the presence of the brane (and presumably of RR flux) might not be
easy. In \cite{ACV} it was found that diffractive string effects may be
relevant even at considerably large impact parameters. This would add
new regimes to the diagram in Figure \ref{bvsE}.

There are a number of sources of other corrections that we have entirely
ignored, such as those due to exchange of particles other than the
graviton, or the finite rest mass of the scattered particles.
Furthermore, any effects due to finite brane thickness have also been
neglected. Again, if the brane thickness is on the scale of the
fundamental length $M_*^{-1}$, the regimes we have considered are not
able to resolve it.

Finally, we made some mention in the introduction to works where the
gravitational scattering at high energies has been studied for its
possible relevance to the problem of extreme energy cosmic rays
\cite{cosmic}. In most of these works the scattering has been considered
in the Born approximation, on the basis that at the relevant energies
gravity is presumably not strongly coupled. We shall not enter at this
stage into the discussion of how to correctly compute the scattering for
the relevant process, and how to account for unitarity. Nevertheless, it
appears like the phenomenological possibility and consequences of
TeV-mass black holes forming in cosmic ray collisions are still to be
developed. The total cross section is presumably dominated by other
softer processes, but still the consequences might be interesting. At
high enough energies one should only need classical general relativity
to describe the process: other interactions and quantum effects will
remain hidden behind the horizon.

\section*{Acknowledgements}

This work was prompted by a conversation with Alex Feinstein, which I
gratefully acknowledge. I would also like to thank Manuel Masip for
conversations. Partial support from UPV grant 063.310-EB187/98 and
CICYT AEN99-0315 is acknowledged.

\end{document}